\documentstyle[preprint,aps]{revtex}
\tighten
\draft
\widetext
\preprint{YITP-01-07}
\bigskip
\bigskip
\begin{document}
\title{A Solitonic 3-Brane in 6D Bulk}
\medskip
\author{Olindo Corradini\footnote{E-mail:
olindo@insti.physics.sunysb.edu} and Zurab Kakushadze\footnote{E-mail: 
zurab@insti.physics.sunysb.edu}}
\bigskip
\address{C.N. Yang Institute for Theoretical Physics\\ 
State University of New York, Stony Brook, NY 11794}

\date{March 5, 2001}
\bigskip
\medskip
\maketitle

\begin{abstract} 
{}We construct a solitonic 3-brane solution in the 6-dimensional
Einstein-Hilbert-Gauss-Bonnet theory with a (negative)
cosmological term. This solitonic brane world is
$\delta$-function-like. Near the brane the metric is that for a product
of the 4-dimensional flat Minkowski space with a 2-dimensional ``wedge''
with a deficit angle (which depends on the solitonic brane tension).
Far from the brane the metric approaches that for a product of the
5-dimensional AdS space and a circle. This solitonic solution
exists for a special value of the Gauss-Bonnet coupling (for which we also
have a $\delta$-function-like codimension-1 solitonic solution), and the
solitonic brane tension can take values in a continuous range. We discuss 
various properties of this solitonic brane world, including coupling between
gravity and matter localized on the brane. 
\end{abstract}

\section{Introduction}

{}In the Brane World scenario the Standard Model gauge and matter fields
are assumed to be localized on  
branes (or an intersection thereof), while gravity lives in a larger
dimensional bulk of space-time 
\cite{early,BK,polchi,witt,lyk,shif,TeV,dienes,3gen,anto,ST,BW,Gog,RS,DGP,DG}. 
There is a big difference between the footings on which
gauge plus matter fields and gravity come in this picture\footnote{This, at
least in some sense, might not be an unwelcome feature - see, {\em e.g.},
\cite{witt,TeV,BW}.}. Thus, for instance, if gauge and matter fields are
localized on D-branes \cite{polchi}, they propagate only in the directions 
along the D-brane world-volume. Gravity, however, is generically 
not confined to the branes - even if we have a graviton zero mode localized
on the brane as in \cite{RS}, where the volume of the extra dimension is 
finite, massive graviton modes are still free to propagate in the bulk. 
On the other hand, as was discussed in \cite{DG}, 
in the cases with infinite volume extra dimensions, we can 
have almost completely localized gravity on higher codimension 
($\delta$-function-like) branes with the $p^2=0$ modes penetrating into the 
bulk.

{}Recently in \cite{alberto} it was pointed out that we can have
{\em complete} localization of gravity on a $\delta$-function-like
{\em solitonic} codimension-1 brane world solution. That is, there are
no propagating degrees of freedom in the bulk, while on the
brane we have 4-dimensional Einstein-Hilbert gravity
(assuming that the solitonic brane is a 3-brane). 
In fact, in this solution, even though the classical solitonic 
background is 5-dimensional, 
the quantum theory perturbatively\footnote{Non-perturbatively 
at the semi-classical level we can {\em a priori} have breakdown of 
causality via creation of ``baby'' branes.} 
is actually 4-dimensional - there are no loop corrections in the bulk
as we have no propagating bulk degrees of freedom.

{}The setup of \cite{alberto}
is the 5-dimensional Einstein-Hilbert theory with a (negative) cosmological
term augmented with a Gauss-Bonnet term. The solitonic brane world
solution arises in this theory for a special value of the Gauss-Bonnet
coupling. The fact that there are no propagating degrees of freedom in the
bulk is then due to a perfect cancellation between the 
corresponding contributions coming from the Einstein-Hilbert and
Gauss-Bonnet terms, which occurs precisely for this value of the
Gauss-Bonnet coupling. Since the bulk theory does not receive loop 
corrections, the classical choice of parameters such as the special value of 
the Gauss-Bonnet coupling (or the Gauss-Bonnet combination itself) 
does {\em not} require perturbative order-by-order fine-tuning. Also, the 
entire setup can be supersymmetrized, and then the aforementioned solitonic 
solution becomes a BPS state, which preserves 1/2 of the original 
supersymmetries.
  
{}In this paper we would like to address the question whether there are higher
codimension solitonic brane world solutions in 
(the appropriate higher dimensional versions of) 
the setup of \cite{alberto}. In fact, we do find codimension-2 {\em 
solitonic}\footnote{Codimension-2 solutions in the 6-dimensional 
Einstein-Hilbert gravity in the presence of source (that is, non-solitonic)
branes were discussed in \cite{CP,GS}.} 
solutions, which are 3-branes if the bulk is 6-dimensional. 
Thus, we have a $\delta$-function-like codimension-2 solitonic solution.
This solution, where the solitonic brane world-volume is flat, exists for a 
continuous range of values of the solitonic brane tension. 
However, as we explain in
the following, this is not a ``self-tuning'' solution for two reasons. First,
it turns out that to have a consistent tree-level coupling between gravity
and brane matter the latter must be conformal.
Second, the aforementioned special choice of 
the Gauss-Bonnet coupling (unlike in the codimension-1 solution of 
\cite{alberto}) is sensitive to quantum corrections in the bulk. This is 
because in this codimension-2 solution we do have propagating degrees of 
freedom in the bulk.

{}The remainder of this paper is organized as follows. In section II we discuss
our setup. In section III we find the aforementioned solitonic codimension-2
brane world solutions and discuss their properties. In section IV we discuss
the coupling between gravity and brane matter. Section V contains various 
remarks. 

\section{The Setup}

{}In this section we discuss the setup within which we will construct the
aforementioned codimension-2 solitonic brane world solutions. 
The action for this model is given by 
(for calculational convenience we will keep the number of space-time 
dimensions $D$ unspecified, but we are mostly interested in the $D=6$ case):
\begin{equation}\label{actionGB}
 S=M_P^{D-2}
 \int d^D x \sqrt{-G} \left\{R+\lambda\left[R^2-4R_{MN}^2+R_{MNST}^2\right]
 -\Lambda\right\}~,
\end{equation}
where $M_P$ is the $D$-dimensional (reduced) Planck scale, and 
the Gauss-Bonnet coupling $\lambda$ has dimension $({\rm length})^2$.
Finally, the bulk vacuum energy density $\Lambda$ is a constant.

{}The equations of motion following form the action (\ref{actionGB}) read:
\begin{eqnarray}
 &&R_{MN}-{1\over 2}G_{MN} R
 -{1\over 2}\lambda G_{MN}\left(R^2-4R^{MN}R_{MN}+R^{MNRS}R_{MNRS}
 \right)+\nonumber\\
 \label{einstein}
 &&2\lambda\left(R R_{MN}-2R_{MS}{R^S}_N+R_{MRST}{R_N}^{RST}-2R^{RS}R_{MRNS}
 \right)+{1\over 2}G_{MN} \Lambda=0~.
\end{eqnarray}
Note that this equation does not contain terms with third and fourth 
derivatives of the metric, which is a special property of the Gauss-Bonnet
combination \cite{Zwiebach,Zumino}.

\subsection{Codimension-1 Solitonic Brane-World}

{}In \cite{alberto} it was shown that, for a special combination of the
Gauss-Bonnet coupling $\lambda$ and the vacuum energy density $\Lambda$,
this theory possesses a codimension-1 {\em solitonic} brane-world solution.
Since this solution will be relevant for our subsequent discussions, let us
briefly review it here. Thus, let us focus on solutions to the above 
equations of motion with the warped \cite{Visser} metric of the form 
\begin{equation}\label{warped}
 ds_D^2=\exp(2A)\eta_{MN} dx^M dx^N~,
\end{equation}
where $\eta_{MN}$ is the flat $D$-dimensional Minkowski metric, and
the warp factor $A$, which is a function of $z\equiv x^D$, 
is independent of the other $(D-1)$ coordinates $x^\mu$.
With this ans{\"a}tz, we have the following
equations of motion for $A$ (prime denotes derivative w.r.t. $z$):
\begin{eqnarray}
 \label{A'GB}
 &&(D-1)(D-2)(A^\prime)^2\left[1-(D-3)(D-4)\lambda(A^\prime)^2
 \exp(-2A)\right]+ \Lambda \exp(2A)=0~,\\
 \label{A''GB}
 &&(D-2)\left[A^{\prime\prime}-(A^\prime)^2\right]
 \left[1-2(D-3)(D-4)\lambda(A^\prime)^2
 \exp(-2A)\right]
 =0~.
\end{eqnarray}
This system of equations has a set of solutions where the $D$-dimensional
space is an AdS space for a continuous range of parameters $\Lambda$ and 
$\lambda$. The volume of the $z$ direction for this set of solutions is
infinite.

{}There, however, also exists a solution where the volume of the
$z$ direction is finite if we ``fine-tune'' the Gauss-Bonnet coupling 
$\lambda$ and the bulk vacuum energy density $\Lambda$ as 
follows\footnote{This special value of the Gauss-Bonnet coupling has appeared
in a somewhat different context in \cite{Zanelli}. In fact, it was argued
in \cite{Zanelli} that for other values of these parameters the 
Einstein-Hilbert-Gauss-Bonnet theory is non-unitary.}:
\begin{equation}\label{fine}
 \Lambda=- {{(D-1)(D-2)}\over{(D-3)(D-4)}}{1\over 4\lambda}~,
\end{equation}
where $\lambda>0$, and $\Lambda<0$. 
This solution is given by (we have chosen the 
integration constant such that $A(0)=0$):
\begin{equation}\label{solutionGB}
 A(z)=-\ln\left[{|z|\over\Delta}+1\right]~,
\end{equation}
where $\Delta$ is given by
\begin{equation}\label{Delta}
 \Delta^2= 2(D-3)(D-4)\lambda~.
\end{equation}
Note that $\Delta$ can be positive or negative. In the former case the
volume of the $z$ direction is finite: $v=2\Delta/(D-1)$. 
On the other hand, in the latter case it is infinite.

{}Note that $A^\prime$ is discontinuous at $z=0$, and $A^{\prime\prime}$
has a $\delta$-function-like behavior at $z=0$. Note, however, that
(\ref{A''GB}) is still satisfied as in this solution
\begin{equation}\label{factor}
 1-2(D-3)(D-4)\lambda(A^\prime)^2\exp(-2A)=0~.
\end{equation}
Thus, this solution
describes a codimension-1 {\em soliton}. The tension of this soliton,
which is given by
\begin{equation}
 f_{D-1}={4(D-2)\over \Delta}M_P^{D-2}~,
\end{equation}
is positive for $\Delta>0$, and it is negative for $\Delta<0$. As was shown in
\cite{alberto}, in the latter case the theory is non-unitary (which is
attributed to the negativity of the brane tension). The solution with positive
brane tension, on the other hand, is consistent. Here we are referring to the 
$z=0$ hypersurface as the brane.

{}It was further shown in \cite{alberto} that the graviton propagator in the 
above solitonic solution vanishes in the bulk, while on the brane we have {\em 
completely} localized gravity. In particular, (at least 
perturbatively\footnote{As was pointed out in \cite{alberto}, 
{\em a priori} semi-classically
there can be non-perturbative effects breaking causality via creation of
``baby'' branes, so that gravity could in this way propagate into the bulk.}) 
gravity on the brane is purely $(D-1)$-dimensional. 

\section{Codimension-2 Solitonic Brane-World}

{}In this section we would like to point out that in the above setup,
precisely for the special combination of the parameters (\ref{fine}), there
also exists a codimension-2 solitonic brane world solution. Thus, consider
the following ans{\"a}tz for the metric:
\begin{equation}
 ds_D^2=\exp(2A)\left[\eta_{\alpha\beta}dx^\alpha dx^\beta +(dr)^2 +
 \exp(2B)~r^2 (d\phi)^2\right]~,
\end{equation}
where $\eta_{\alpha\beta}$ is the flat $(D-2)$-dimensional Minkowski metric
corresponding to the first $(D-2)$ coordinates $x^\alpha$, and the other two
coordinates are chosen in the polar basis $(r,\phi)$; the warp factors
$A$ and $B$, which are functions of $r$, are assumed to be independent of 
$(x^\alpha,\phi)$ (that is, we are looking for axially symmetric 
solutions); the
angular coordinate $\phi$ takes 
values between 0 and $2\pi$, while the radial
coordinate $r$ takes values between $0$ and $\infty$.

{}With the above ans{\"a}tz we have the following equations of motion for
$A$ and $B$ (dot denotes derivative w.r.t. $r$):
\begin{eqnarray}
 &&\left[\ddot B+{2\over r}~\dot B+ {1\over r^2}~
 \dot B^2+(D-2)\left(\ddot A-\dot A^2\right)\right]
 \Biggl[1-2(D-3)(D-4)\lambda\dot A^2
 \exp(-2A)\Biggr]\nonumber\\
 \label{eq1}
 &&-4(D-3)(D-4)\lambda\left[\ddot A-\dot A^2\right]\dot A\left[\dot B+{1\over
 r}\right] \exp(-2A)
 =0~,\\
 &&(D-1)(D-2)\dot A^2\Biggl[1-(D-3)(D-4)\lambda\dot A^2\exp(-2A)\Biggr]
 +\Lambda \exp(2A)\nonumber\\
 \label{eq2}
 &&+2(D-2)\dot A\left[\dot B+{1\over r}\right]
 \Biggl[1-2(D-3)(D-4)\lambda \dot A^2 
 \exp(-2A)\Biggr]=0~,\\
 &&(D-1)(D-2)\dot A^2\Biggl[1-(D-3)(D-4)\lambda \dot A^2\exp(-2A)\Biggr]
 + \Lambda\exp(2A)\nonumber\\ 
 \label{eq3}
 &&+2(D-2)\left[\ddot A-\dot A^2\right]\Biggl[1-
 2(D-3)(D-4)\lambda\dot A^2\exp(-2A)\Biggr]=0~.
\end{eqnarray}
The first equation is a linear combination of the $(\alpha\beta)$ and
$(rr)$ equations, the second equation is the $(rr)$ equation, and the
third equation is the $(\phi\phi)$ equation. Only two of the above three
equations are independent, which, as usual, is a consequence of Bianchi
identities.

\subsection{$\delta$-function-like Solitonic Brane World}

{}Here we would like to discuss a solution of the above equations of motion
corresponding to a $\delta$-function-like codimension-2 solitonic brane world.
This solution is given by:
\begin{equation}\label{delta}
 A(r)=-\ln\left({r\over \Delta}+1\right)~,~~~
 B(r)=-\beta~,
\end{equation}
where $\beta$ is a constant, which we will assume to be {\em positive}, 
while $\Delta$, which is also assumed to be positive, 
is related to $\lambda$ via (\ref{Delta}).
The metric is given by:
\begin{equation}\label{metric0}
 ds_D^2=\left({r\over \Delta}+1\right)^{-2}
 \left[\eta_{\alpha\beta} dx^\alpha dx^\beta+(dr)^2+\exp(-2\beta)~r^2 
 (d\phi)^2\right]~. 
\end{equation} 
Note that $\Delta\rightarrow \infty$ in this solution corresponds to the flat
bulk limit. Also note that the presence of the Gauss-Bonnet term (as well as
the fact that the Gauss-Bonnet coupling takes a special value (\ref{fine}))
is crucial for the existence of the solution (\ref{delta}) - indeed, 
(\ref{eq2}) could not be satisfied without the Gauss-Bonnet term.

{}Near the origin $r\rightarrow 0$ the above metric takes the following form:
\begin{equation}\label{metric01}
 ds_D^2=\eta_{\alpha\beta} dx^\alpha dx^\beta+(dr)^2+\exp(-2\beta)~r^2 
 (d\phi)^2~. 
\end{equation}
This metric describes a product of the $(D-2)$-dimensional flat Minkowski
space with a 2-dimensional ``wedge'' with the deficit angle
\begin{equation}\label{theta}
 \theta=2\pi\left[1-\exp(-\beta)\right]~.
\end{equation}
This wedge is locally ${\bf R}^2$ except for the origin $r=0$, where we have 
a $\delta$-function-like singularity. Thus, we have a  
$\delta$-function-like codimension-2 {\em solitonic} brane located at $r=0$. 

{}The tension of this solitonic brane can be determined as follows. 
Consider the following action
\begin{equation}\label{action1}
 S_1=M_P^{D-2}\int d^D x\sqrt{-G}~R-f_{D-2}\int_\Sigma d^{D-2}x
 \sqrt{-{\widehat G}}~,
\end{equation}
where $\Sigma$ is a $\delta$-function-like codimension-2 {\em source} brane,
which is the hypersurface $x^i=0$ ($x^i$, $i=1,2$, are the two coordinates 
transverse to the brane); the tension $f_{D-2}$ of this brane is assumed to be
positive; finally, 
\begin{equation}
 {\widehat G}_{\alpha\beta}\equiv{\delta_\alpha}^M{\delta_\beta}^N G_{MN}
 \Big|_\Sigma~,
\end{equation} 
where $x^\alpha$ are the $(D-2)$ coordinates along the brane (the 
$D$-dimensional coordinates are given by $x^M=(x^\alpha,x^i)$).

{}The equations of motion following from the action (\ref{action1}) are
given by:
\begin{equation}
 R_{MN}-{1\over 2}G_{MN} R+{1\over 2}{\sqrt{-{\widehat G}}\over
 \sqrt{-G}}~{\delta_M}^\alpha {\delta_N}^\beta {\widehat G}_{\alpha\beta}~
 {\widetilde f}_{D-2}~\delta^{(2)}(x^i)=0~,
\end{equation}
where ${\widetilde f}_{D-2}\equiv f_{D-2}/M_P^{D-2}$.

{}Next, consider the following ans{\"a}tz for the metric:
\begin{equation}
 ds_D^2=\eta_{\alpha\beta}dx^\alpha dx^\beta+
 \exp(2\omega)~\delta_{ij} dx^i dx^j~,
\end{equation}
where $\omega$ is a function of $x^i$ but is independent of $x^\alpha$.
With this ans{\"a}tz we have
\begin{eqnarray}
 &&R_{\alpha\beta}=0~,~~~
 R_{ij}={\widetilde R}_{ij}={1\over 2} {\widetilde G}_{ij}
 {\widetilde R}~,\\
 &&\sqrt{\widetilde G}~{\widetilde R}={\widetilde f}_{D-2}~\delta^{(2)}(x^i)~,
\end{eqnarray} 
where ${\widetilde R}$ and ${\widetilde R}_{ij}$ are the 2-dimensional
Ricci scalar respectively Ricci tensor constructed from the 2-dimensional
metric 
\begin{equation}
 {\widetilde G}_{ij}\equiv \exp(2\omega)~\delta_{ij}~. 
\end{equation}
Since this metric is conformally flat, we have $\sqrt{\widetilde G}
{\widetilde R}=-2\partial^i\partial_i\omega$ (where the indices are lowered 
and raised using $\delta_{ij}$ and $\delta^{ij}$, respectively), so that
\begin{equation}
 \partial^i\partial_i\omega=-{1\over 2}{\widetilde f}_{D-2}~\delta^{(2)}(x^i)~.
\end{equation}
The solution to this equation is given by
\begin{equation}
 \omega=-{1\over 8\pi}{\widetilde f}_{D-2}\ln\left(x^2\over a^2\right)~,
\end{equation}
where $x^2\equiv \delta_{ij}x^ix^j$~, and $a$ is an integration constant.
 
{}Let us go to the polar coordinates $(\rho,\phi)$: $x^1=\rho\cos(\phi)$,
$x^2=\rho\sin(\phi)$ ($\rho$ takes values from 0 to $\infty$, while $\phi$
takes values from 0 to $2\pi$). In these coordinates the two dimensional
metric is given by
\begin{equation}
 d{\widetilde s}_2^2=\left({a^2\over \rho^2}\right)^\nu\left[(d\rho)^2+
 \rho^2(d\phi)^2\right]~,
\end{equation}
where
\begin{equation}
 \nu\equiv{1\over 4\pi}{\widetilde f}_{D-2}~.
\end{equation}
Let us change the coordinates to $(r,\phi)$, where 
\begin{equation}
 r\equiv{1\over {1-\nu}}~a^\nu\rho^{1-\nu}~,
\end{equation}
where we are assuming that $\nu<1$. Then we have
\begin{equation}
 d{\widetilde s}_2^2=(dr)^2+
 \exp(-2\beta)~r^2(d\phi)^2~,
\end{equation}
where
\begin{equation}
 \exp(-\beta)\equiv 1-\nu~.
\end{equation}
Thus, we see that the brane tension $f_{D-2}$ is related to the deficit angle
$\theta$ (given by (\ref{theta})) via
\begin{equation}
 f_{D-2}=2 M_P^{D-2}~\theta~.
\end{equation}
In particular, this expression gives the tension of the 
$\delta$-function-like codimension-2 solitonic brane located at $r=0$ in the
solution described by the metric (\ref{metric0}).

{}Before we end this section let us note that for large $r$ ($r\gg \Delta$)
the metric (\ref{metric0}) approaches that of AdS$_{D-1}\times S^1$
\begin{equation}\label{metric02}
 ds_D^2={\Delta^2\over r^2}\left[\eta_{\alpha\beta}dx^\alpha dx^\beta+
 (dr)^2\right]+r_*^2 (d\phi)^2~,
\end{equation}
where the radius of $S^1$ is given by $r_*\equiv\Delta\exp(-\beta)$.

\section{Coupling to Brane Matter}

{}In this section we would like to discuss how gravity couples to
matter localized on the above $\delta$-function-like codimension-2 solitonic 
brane. Since this brane is solitonic, it breaks $D$-dimensional 
diffeomorphisms only spontaneously. This has certain implications to which we
now turn.

{}Thus, let us consider small fluctuations around the solution (\ref{metric0}):
\begin{equation}
 G_{MN}=\exp(2A)\left[{\overline G}_{MN}+{\overline h}_{MN}\right]~,
\end{equation}
where ${\overline G}_{MN}$ is the background metric up to the warp factor
$\exp(2A)$ (that is, ${\overline G}_{\alpha\beta}=\eta_{\alpha\beta}$,
${\overline G}_{rr}=1$, ${\overline G}_{\phi\phi}=\exp(-2\beta)r^2$,
${\overline G}_{\alpha r}={\overline G}_{\alpha\phi}={\overline G}_{r\phi}=0$),
and for convenience reasons instead of the metric fluctuations $h_{MN}\equiv
\exp(2A){\overline h}_{MN}$ we choose to work with ${\overline h}_{MN}$.

{}In terms of ${\overline h}_{MN}$ the $D$-dimensional diffeomorphisms 
(corresponding to $x^M\rightarrow x^M-\xi^M$) read:
\begin{eqnarray}
 &&\delta{\overline h}_{\alpha\beta}=
 \partial_\alpha {\overline\xi}_\beta+\partial_\beta{\overline\xi}_\alpha+
 2\eta_{\alpha\beta}{\dot A}{\overline\xi}_r~,\\
 &&\delta {\overline h}_{\alpha r}=\partial_\alpha{\overline\xi}_r+
 {\dot{\overline\xi}}_\alpha~,\\
 &&\delta{\overline h}_{\alpha\phi}=\partial_\alpha{\overline\xi}_\phi+
 \partial_\phi{\overline \xi}_\alpha~,\\
 &&\delta{\overline h}_{rr}=2{\dot{\overline \xi}}_r+2{\dot A}
 {\overline \xi}_r~,\\
 &&\delta{\overline h}_{r\phi}={\dot{\overline\xi}}_\phi-
 {2\over r}{\overline \xi}_\phi+\partial_\phi{\overline\xi}_r~,\\
 &&\delta{\overline h}_{\phi\phi}=2\partial_\phi{\overline\xi}_\phi
 +2\exp(-2\beta)r^2\left[{\dot A}+{1\over r}\right]{\overline\xi}_r~,
\end{eqnarray}
where ${\overline \xi}_M\equiv\exp(-2A)\xi_M$. Note that using these
diffeomorphisms we can set two of the graviscalar components (${\overline
h}_{rr}$ and ${\overline h}_{r\phi}$) as well as one of the graviphotons
(${\overline h}_{\alpha r}$) to zero. We are then left with the 
$(D-2)$-dimensional graviton (${\overline h}_{\alpha\beta}$), a graviphoton
(${\overline h}_{\alpha\phi}$), and a graviscalar (${\overline h}_{\phi\phi}$).

{}Next, let us assume that we have matter localized on the 
$\delta$-function-like codimension-2 solitonic brane. Let $T_{\alpha\beta}$
be the corresponding conserved energy momentum tensor:
\begin{equation}
 \partial^\alpha T_{\alpha\beta}=0~.
\end{equation}
The coupling of gravity to the brane matter is described by the following
term in the action:
\begin{equation}
 {1\over 2}\int_\Sigma d^{D-2}x~T^{\alpha\beta}{\overline h}_{\alpha\beta}~,
\end{equation}
where $\Sigma$ is the $r=0$ hypersurface corresponding to the brane
(note that on $\Sigma$ $h_{\alpha\beta}$ and ${\overline h}_{\alpha\beta}$
coincide as $A(r=0)=0$). Since ${\dot A}(r=0)=-1/\Delta\not=0$, this coupling
is invariant under the aforementioned diffeomorphisms if and only if
\begin{equation}
 T\equiv T_\alpha^\alpha=0~,
\end{equation}
that is, if and only if the brane matter is {\em conformal}.

{}So, at the tree level, to have a consistent coupling between gravity and the
brane matter we must assume that the latter is conformal\footnote{This is
similar to what happens in the setup of \cite{zura}.}. Note, however, that
the conformal property cannot generically persist beyond the tree-level. 
Indeed, the volume of the extra two dimensions in the above solution is finite:
\begin{equation}
 {\widetilde v}_2=\int_0^\infty dr\int_0^{2\pi} d\phi\sqrt{-{\overline G}}
 \exp(DA)=2\pi\exp(-\beta){\Delta^2\over(D-1)(D-2)}~.
\end{equation} 
This implies that the $(D-2)$-dimensional Planck scale on the brane is 
finite:
\begin{equation}
 {\widehat M}_P^{D-4}=M_P^{D-2}
 \int_0^\infty dr\int_0^{2\pi} d\phi\sqrt{-{\overline G}}
 \exp[(D-2)A]=4\pi\lambda\exp(-\beta)M_P^{D-2}~.
\end{equation}
That is, we have a quadratically normalizable $(D-2)$-dimensional graviton 
zero mode. The conformal invariance of the matter localized on the brane is 
then generically expected to be broken by loop corrections involving gravity.

{}Thus, in the above brane world solution we have a quantum inconsistency
discussed in \cite{COSM} in other setups. Note that such an inconsistency
does not arise in the setup of \cite{alberto}. The key reason is that in
the codimension-1 solution of \cite{alberto} (which we reviewed in
section II) there are no propagating degrees of freedom in the bulk.
In the above codimension-2 solution, however, we do have such degrees of
freedom, in particular, the aforementioned graviscalar degree of freedom
${\overline h}_{\phi\phi}$ propagates in the bulk. A simple way to see 
this is to recall that, as we pointed out at the end of the previous section,
at large $r$ the metric in this solution approaches that of AdS$_{D-1}\times
S^1$, so we do have a propagating degree of freedom corresponding to the
reduction on $S^1$. This then implies that (unlike in the codimension-1 
solution) here we do have loop corrections in the bulk, and the corresponding
higher curvature terms will generically delocalize gravity \cite{COSM}.

\section {Comments}

{}We would like to end our discussion with a few remarks. First, note that
the above codimension-2 solution, where  the brane world-volume is flat,
exists for a continuous range of values of the solitonic brane 
tension\footnote{More precisely, it exists for 
$0<f_{D-2}<f_{\rm{\small crit}}$; at the critical brane tension
$f_{\rm{\small crit}}=4\pi M_P^{D-2}$ the deficit angle $\theta=2\pi$.}.
However, this is not a ``self-tuning'' solution for two reasons. First,
to have a consistent tree-level coupling between gravity
and brane matter the latter must be conformal. 
Second, the aforementioned special choice of 
the Gauss-Bonnet coupling (unlike in the codimension-1 solution of 
\cite{alberto}) is sensitive to quantum corrections in the bulk. 

{}Note that the issues (which are expected to arise at 
the quantum level as we discussed at the end of the previous section) 
with the coupling between
the brane matter and gravity as well as with delocalization of gravity by 
higher curvature terms in the bulk need not arise in scenarios with
infinite volume extra dimensions \cite{GRS,CEH,DGP0,witten,DVALI}, 
at least in higher
codimension cases. Here the four-dimensional gravity on the brane
is obtained via the Einstein-Hilbert term on the brane, which is generically
expected to be generated by quantum effects on the brane as long as the
brane world-volume theory is not conformal \cite{DGP,DG}. As was pointed out
in \cite{alberto}, this is expected to be the case in the string theory
context as well. 
Thus, in the orbifold examples of \cite{KS,LNV,BKV}\footnote{The 
following is also correct for the orientifold examples of \cite{orient}.
In considering such examples with, say, ${\cal N}=1$ supersymmetry, however,
some caution is needed due to the subtleties discussed in \cite{KaSh,KST}.}
we always have non-conformal $U(1)$ factors. Also, in other examples such as
conifolds \cite{Kehagias,KW} already the non-Abelian gauge subgroups are
non-conformal in the ultra-violet (albeit they are conformal in the infra-red).
As was argued in \cite{alberto}, in the aforementioned examples (which are
conformal in the infra-red, but are non-conformal in the ultra-violet) 
in the context of AdS/CFT correspondence \cite{malda,GKP,WITT} on the
Type IIB side various higher curvature terms intrinsically due to the 
compactification\footnote{These should not be confused with the higher 
dimensional terms already present in ten dimensions, which do not affect
conformality \cite{Banks,Kallosh,Gutperle}.} become important. Finally, 
recently non-conformal theories (that is, theories that are not conformal 
even in the infra-red) were discussed in \cite{radu} within a modification of 
the setup of \cite{BKV}. Some of these theories can be discussed in the context
of a certain brane-bulk duality \cite{radu}, which might provide a framework
for computing gravitational corrections (such as the Einstein-Hilbert term)
on D-branes. 

\acknowledgments

{}We would like to thank Gregory Gabadadze and Slava Zhukov for
valuable discussions.
This work was supported in part by the National Science Foundation.
Z.K. would also like to thank Albert and Ribena Yu for financial support.


\begin{references}

\bibitem{early} 
V. Rubakov and M. Shaposhnikov, Phys. Lett. {\bf B125} (1983) 136.

\bibitem{BK}
A. Barnaveli and O. Kancheli, Sov. J. Nucl. Phys. {\bf 52} (1990) 576.

\bibitem{polchi} J. Polchinski, Phys. Rev. Lett. {\bf 75} (1995) 4724.

\bibitem{witt} P. Ho{\u r}ava and E. Witten, Nucl. Phys. {\bf B460} (1996)
506; Nucl. Phys. {\bf B475} (1996) 94;\\
E. Witten, Nucl. Phys. {\bf B471} (1996) 135.

\bibitem{lyk} I. Antoniadis, Phys. Lett. {\bf B246} (1990) 377;\\
J. Lykken, Phys. Rev. {\bf D54} (1996) 3693.

\bibitem{shif} G. Dvali and M. Shifman, Nucl. Phys. {\bf B504} (1997) 127;
Phys. Lett. {\bf B396} (1997) 64.

\bibitem{TeV} N. Arkani-Hamed, S. Dimopoulos and G. Dvali, 
Phys. Lett. {\bf B429} (1998) 263; Phys. Rev. {\bf D59} (1999) 086004.

\bibitem{dienes} K.R. Dienes, E. Dudas and T. Gherghetta, Phys. Lett. 
{\bf B436} (1998) 55; Nucl. Phys. {\bf B537} (1999) 47; hep-ph/9807522;\\
Z. Kakushadze, Nucl. Phys. {\bf B548} (1999) 205; Nucl. Phys.
{\bf B552} (1999) 3;\\
Z. Kakushadze and T.R. Taylor, Nucl. Phys. {\bf B562} (1999) 78.

\bibitem{3gen} Z. Kakushadze, Phys. Lett. {\bf B434} (1998) 269; 
Nucl. Phys. {\bf B535} (1998) 311; Phys. Rev. {\bf D58} (1998) 101901.

\bibitem{anto} I. Antoniadis, N. Arkani-Hamed, S. Dimopoulos and G. Dvali,
Phys. Lett. {\bf B436} (1998) 257.

\bibitem{ST} G. Shiu and S.-H.H. Tye, Phys. Rev. {\bf D58} (1998) 106007.

\bibitem{BW} Z. Kakushadze and S.-H.H. Tye, Nucl. Phys. {\bf B548} (1999) 180;
Phys. Rev. {\bf D58} (1998) 126001.

\bibitem{Gog} M. Gogberashvili, hep-ph/9812296; Europhys. Lett. {\bf 49} 
(2000) 396.

\bibitem{RS} L. Randall and R. Sundrum, Phys. Rev. Lett. {\bf 83} (1999)
3370; Phys. Rev. Lett. {\bf 83} (1999) 4690.

\bibitem{DGP} G. Dvali, G. Gabadadze and M. Porrati, Phys. Lett. {\bf 
B485} (2000) 208.

\bibitem{DG} G. Dvali and G. Gabadadze, Phys. Rev. {\bf D63} (2001) 065007.

\bibitem{alberto} A. Iglesias and Z. Kakushadze, hep-th/0011111; 
hep-th/0012049.

\bibitem{CP} A. Chodos and E. Poppitz, Phys. Lett. {\bf B471} (1999) 119.

\bibitem{GS} T. Gherghetta and M. Shaposhnikov, Phys. Rev. Lett.
{\bf 85} (2000) 240.

\bibitem{Zwiebach} B. Zwiebach, Phys. Lett. {\bf B156} (1985) 315.

\bibitem{Zumino} B. Zumino, Phys. Rept. {\bf 137} (1986) 109.

\bibitem{Visser} M. Visser, Phys. Lett. {\bf B159} (1985) 22;\\
P. van Nieuwenhuizen and N.P. Warner, Commun. Math. Phys. {\bf 99} (1985)
141.

\bibitem{Zanelli} J. Cris{\'o}stomo, R. Troncoso and J. Zanelli, 
Phys. Rev. {\bf D62} (2000) 084013.

\bibitem{zura} Z. Kakushadze, Phys. Lett. {\bf B488} (2000) 402;
Phys. Lett. {\bf B489} (2000) 207; Phys. Lett. {\bf B491} (2000) 317;
Mod. Phys. Lett. {\bf A15} (2000) 1879.

\bibitem{COSM} Z. Kakushadze, Nucl. Phys. {\bf B589} (2000) 75;
Phys. Lett. {\bf B497} (2001) 125;\\
O. Corradini and Z. Kakushadze, Phys. Lett. {\bf B494} (2000) 302;\\
Z. Kakushadze and P. Langfelder, Mod. Phys. Lett. {\bf A15} (2000) 2265.

\bibitem{GRS} R. Gregory, V.A. Rubakov and S.M. Sibiryakov, 
Phys. Rev. Lett. {\bf 84} (2000) 5928.

\bibitem{CEH} C. Csaki, J. Erlich and T.J. Hollowood, Phys. Rev. Lett. {\bf
84} (2000) 5932.

\bibitem{DGP0} G. Dvali, G. Gabadadze and M. Porrati, Phys. Lett. {\bf B484}
(2000) 112; Phys. Lett. {\bf B484} (2000) 129.

\bibitem{witten} E. Witten, hep-ph/0002297.

\bibitem{DVALI} G. Dvali, hep-th/0004057.

\bibitem{KS} S. Kachru and E. Silverstein, Phys. Rev. Lett. {\bf 80}
(1998) 4855.

\bibitem{LNV} A. Lawrence, N. Nekrasov and C. Vafa, 
Nucl. Phys. {\bf B533} (1998) 199.

\bibitem{BKV} 
M. Bershadsky, Z. Kakushadze and C. Vafa, Nucl. Phys. {\bf B523} 
(1998) 59.

\bibitem{orient} 
Z. Kakushadze, Nucl. Phys. {\bf B529} (1998) 157;  Phys. Rev. {\bf D58} 
(1998) 106003; Phys. Rev. {\bf D59} (1999) 045007; Nucl. Phys. {\bf B544} 
(1999) 265.

\bibitem{KaSh} Z. Kakushadze, Nucl. Phys. {\bf B512} (1998) 221;\\
Z. Kakushadze and G. Shiu, Phys. Rev. {\bf D56} (1997) 3686;
Nucl. Phys. {\bf B520} (1998) 75.

\bibitem{KST}Z. Kakushadze, G. Shiu and S.-H.H. Tye,
Nucl. Phys. {\bf B533} (1998) 25;\\ 
Z. Kakushadze, 
Phys. Lett. {\bf B455} (1999) 120; Int. J. Mod. Phys. {\bf A15} (2000) 3461;
Phys. Lett. {\bf B459} (1999) 497.

\bibitem{Kehagias} A. Kehagias, Phys. Lett. {\bf B435} (1998) 337.

\bibitem{KW} I.R. Klebanov and E. Witten, Nucl. Phys. {\bf B536} (1998) 199.

\bibitem{malda} J.M. Maldacena, Adv. Theor. Math. Phys. {\bf 2} (1998) 231.

\bibitem{GKP} S.S. Gubser, I.R. Klebanov and
A.M. Polyakov, Phys. Lett. {\bf B428} (1998) 105.

\bibitem{WITT} E. Witten, Adv. Theor. Math. Phys. 
{\bf 2} (1998) 253.

\bibitem{Banks} T. Banks and M.B. Green, JHEP {\bf 9805} (1998) 002.

\bibitem{Kallosh} R. Kallosh and A. Rajaraman, Phys. Rev. {\bf D58} 
(1998) 125003.  

\bibitem{Gutperle} M.B. Green and M. Gutperle, 
Phys. Rev. {\bf D58} (1998) 046007;\\
M.B. Green, hep-th/9903124, and references therein.

\bibitem{radu} Z. Kakushadze and R. Roiban, hep-th/0102125.

\end{references}
\end{document}